%% file: main.tex
\documentclass[conference]{IEEEtran}
\IEEEoverridecommandlockouts
\usepackage{cite}
\usepackage{amsmath,amssymb,amsfonts,bbm}
\usepackage{mathtools}
\usepackage{graphicx}
\usepackage{textcomp}
\usepackage{xcolor}
\usepackage[acronym]{glossaries}
\usepackage{amsfonts}
\usepackage{siunitx}
\usepackage{booktabs}
\usepackage{subfiles}

\usepackage{algorithm}
\usepackage{algpseudocode}
\usepackage{amsthm}
\usepackage{soul}
\usepackage{multirow}

\input{acronyms}
\DeclareSIUnit{\million}{M}
\DeclareSIUnit\sample{S}
\DeclareSIUnit{\decibelw}{dBW}
\DeclareSIUnit\decibelm{dBm}
\def\BibTeX{{\rm B\kern-.05em{\sc i\kern-.025em b}\kern-.08em
    T\kern-.1667em\lower.7ex\hbox{E}\kern-.125emX}}
\newcommand{\ctext}[3][RGB]{%
  \begingroup
  \definecolor{hlcolor}{#1}{#2}\sethlcolor{hlcolor}%
  \hl{#3}%
  \endgroup
}

\begin{document}

\bstctlcite{IEEEexample:BSTcontrol}

\title{Distributed Learning for Wi-Fi AP Load Prediction}

\author{
\IEEEauthorblockN{Dariush Salami$^{\star}$, Francesc Wilhelmi$^{\flat}$, Lorenzo Galati-Giordano$^{\flat}$, and Mika Kasslin$^{\star}$ \vspace{0.1cm}
} 
\IEEEauthorblockA{$^{\star}$\emph{Radio Systems Research, Nokia Bell Labs, Espoo, Finland}}
\IEEEauthorblockA{$^{\flat}$\emph{Radio Systems Research, Nokia Bell Labs, Stuttgart, Germany}}
}

\maketitle

\subfile{sections/0_abstract}


\subfile{sections/1_introduction}

\subfile{sections/2_related_work}

\subfile{sections/3_distributed_learning}

\subfile{sections/4_methodology}

\subfile{sections/5_results}

\subfile{sections/6_conclusions}

\bibliographystyle{IEEEtran}
\bibliography{references}

\end{document}

%% file: acronyms.tex
\makeglossaries
\newacronym{iot}{IoT}{Internet of Things}
\newacronym{rf}{RF}{Radio Frequency}
\newacronym{hcs}{HCS}{Human-Centric Sensing}
\newacronym{lis}{LIS}{Large Intelligent Surface}
\newacronym{jrc}{JRC}{Joint Radar and Communication}
\newacronym{cw}{CW}{Continuous Wave}
\newacronym{fmcw}{FMCW}{Frequency-Modulated Continuous Wave}
\newacronym{arima}{ARIMA}{Autoregressive Integrated Moving Average}
\newacronym{mape}{MAPE}{Mean Average Percentage Error}
\newacronym{fedavg}{FedAvg}{Federated Averaging}
\newacronym{bs}{BS}{Base Station}
\newacronym{ml}{ML}{Machine Learning}
\newacronym{ann}{ANN}{Artificial Neural Network}
\newacronym{nlp}{NLP}{Natural Language Processing}
\newacronym{mmWave}{mmWave}{Millimeter-Wave}
\newacronym{if}{IF}{Intermediate Frequency}
\newacronym{fft}{FFT}{Fast Fourier Transform}
\newacronym{aoa}{AOA}{Angle of Arrival}
\newacronym{adc}{ADC}{Analog to Digital Converter}
\newacronym{cfar}{CFAR}{Constant False Alarm Rate}
\newacronym{svm}{SVM}{Support Vector Machine}
\newacronym{csi}{CSI}{Channel State Information}
\newacronym{cnn}{CNN}{Convolutional Neural Network}
\newacronym{rnn}{RNN}{Recurrent Neural Network}
\newacronym{iq}{IQ}{In-phase and Quadrature}
\newacronym{rcs}{RCS}{Radar Cross-Section}
\newacronym{ro}{RO}{Reverse Osmosis}
\newacronym{roc}{ROC}{Receiver Operating Characteristic}
\newacronym{auc}{AUC}{Area Under \gls{roc} Curve}
\newacronym{ap}{AP}{Access Point}
\newacronym{sta}{STA}{Station}
\newacronym{hci}{HCI}{Human-Computer Interaction}
\newacronym{ahc}{AHC}{Agglomerative Hierarchical Clustering}
\newacronym{cl}{CL}{Centralized Learning}
\newacronym{lstm}{LSTM}{Long Short-Term Memory}
\newacronym{mlp}{MLP}{Multi-Layer Perceptron}
\newacronym{tfnet}{TFNet}{Transformer Network}
\newacronym{mpnn}{MPNN}{Message Passing Neural Network}
\newacronym{knn}{K-NN}{K-Nearest Neighbor}
\newacronym{gpu}{GPU}{Graphical Processing Unit}
\newacronym{gflop}{GFLOP}{Giga Floating Point OPeration}
\newacronym{eer}{EER}{Equal Error Rate}
\newacronym{rl}{RL}{Reinforcement Learning}
\newacronym{nr}{NR}{New Radio}
\newacronym{mac}{MAC}{Media Access Control}
\newacronym{gmac}{GMAC}{Giga Multiply–Accumulate Operation}
\newacronym{ofdm}{OFDM}{Orthogonal Frequency-Division Multiplexing}
\newacronym{ue}{UE}{User Equipment}
\newacronym{lte}{LTE}{Long Term Evolution}
\newacronym{d2d}{D2D}{Device-To-Device}
\newacronym{enb}{eNB}{eNodeB}
\newacronym{prose}{ProSe}{Proximity Service}
\newacronym{plmn}{PLMN}{Public Land Mobile Network}
\newacronym{3gpp}{3GPP}{3rd Generation Partnership Project}
\newacronym{eutra}{E-UTRA}{Evolved Universal Terrestrial Radio Access}
\newacronym{pssch}{PSSCH}{Physical Sidelink Shared CHannel}
\newacronym{qpsk}{QPSK}{Quadrature Phase Shift Keying}
\newacronym{16qam}{16QAM}{16-state Quadrature Amplitude Modulation}
\newacronym{pdsch}{PDSCH}{Physical Downlink Shared CHannel}
\newacronym{pscch}{PSCCH}{Physical Sidelink Control CHannel}
\newacronym{pdcch}{PDCCH}{Physical Downlink Control CHannel}
\newacronym{psdch}{PSDCH}{Physical Sidelink Discovery CHannel}
\newacronym{psbch}{PSBCH}{Physical Sidelink Broadcast CHannel}
\newacronym{mibsl}{MIB-SL}{Master Information Block-SideLink}
\newacronym{gnb}{gNB}{gNodeB}
\newacronym{rrc}{RRC}{Radio Resource Control}
\newacronym{slrnti}{SL-RNTI}{SideLink Radio Network Temporary Identifier}
\newacronym{dci}{DCI}{Downlink Control Information}
\newacronym{urllc}{URLLC}{Ultra-Reliable and Low-Latency Communication}
\newacronym{5gc}{5GC}{5G Core}
\newacronym{sci}{SCI}{Sidelink Control Information}
\newacronym{cv2x}{C-V2X}{Cellular Vehicle-To-Everything}
\newacronym{prs}{PRS}{Positioning Reference Signal}
\newacronym{srs}{SRS}{Sounding Reference Signal}
\newacronym{v2x}{V2X}{Vehicle-To-Everything}
\newacronym{v2v}{V2V}{Vehicle-To-Vehicle}
\newacronym{scs}{SCS}{Sub-Carrier Spacing}
\newacronym{qos}{QoS}{Quality of Service}
\newacronym{psfch}{PSFCH}{Physical Sidelink Feedback CHannel}
\newacronym{harq}{HARQ}{Hybrid Automatic Repeat Request}
\newacronym{cqi}{CQI}{Channel Quality Information}
\newacronym{cp}{CP}{Cyclic Prefix}
\newacronym{spss}{S-PSS}{Sidelink Primary Synchronization Signal}
\newacronym{ssss}{S-SSS}{Sidelink Secondary Synchronization Signal}
\newacronym{pfrs}{PFRS}{Phase Tracking Reference Signal}
\newacronym{csirs}{CSI-RS}{\gls{csi}-Reference Signal}
\newacronym{cran}{CRAN}{Cloud-Radio Access Network}
\newacronym{fsk}{FSK}{Frequency Shift Keying}
\newacronym{mfsk}{MFSK}{Multiple Frequency-Shift Keying}
\newacronym{ppo}{PPO}{Proximal Policy Optimization Algorithms}
\newacronym{trpo}{TRPO}{Trust Region Policy Optimization}
\newacronym{a2c}{A2C}{Advantage Actor-Critic}
\newacronym{rmse}{RMSE}{Root Mean Square Error}
\newacronym{rss}{RSS}{Received Signal Strength}
\newacronym{dbscan}{DBSCAN}{Density-Based Spatial Clustering of Applications with Noise}
\newacronym{mimo}{MIMO}{Multiple-Input Multiple-Output}
\newacronym{mf}{MF}{Matched Filter}
\newacronym{mse}{MSE}{Mean Squared Error}
\newacronym{iou}{IOU}{Intersection Over Union}
\newacronym{bim}{BIM}{Building Information Modeling}
\newacronym{adas}{ADAS}{Advanced Driver-Assistance System}
\newacronym{pca}{PCA}{Principal Component Analysis}
\newacronym{relu}{ReLU}{Rectified Linear Unit}
\newacronym{imu}{IMU}{Inertial Measurement Unit}
\newacronym{snr}{SNR}{Signal-to-Noise Ratio}
\newacronym{gnn}{GNN}{Graph Neural Network}
\newacronym{ai}{AI}{Artificial Intelligence}
\newacronym{darpa}{DARPA}{Defense Advanced Research Projects Agency}
\newacronym{sdg}{SDG}{Sustainable Development Goal}
\newacronym{un}{UN}{United Nations}
\newacronym{fl}{FL}{Federated Learning}
\newacronym{kd}{KD}{Knowledge Distillation}
\newacronym{iid}{IID}{Independent and Identically Distributed}
\newacronym{kde}{KDE}{Kernel Density Estimation}
\newacronym{gmm}{GMM}{Gaussian Mixture Model}
\newacronym{jsd}{JSD}{Jensen-Shannon Divergence}
\newacronym{kld}{KLD}{Kullback–Leibler Divergence}
\newacronym{mae}{MAE}{Mean Absolute Error}
\glsaddall

%% file: sections/0_abstract.tex
\begin{abstract}
The increasing cloudification and softwarization of networks foster the interplay among multiple independently managed deployments. An appealing reason for such an interplay lies in distributed \gls{ml}, which allows the creation of robust \gls{ml} models by leveraging collective intelligence and computational power. In this paper, we study the application of the two cornerstones of distributed learning, namely \gls{fl} and \gls{kd}, on the Wi-Fi \gls{ap} load prediction use case. The analysis conducted in this paper is done on a dataset that contains real measurements from a large Wi-Fi campus network, which we use to train the \gls{ml} model under study based on different strategies. Performance evaluation includes relevant aspects for the suitability of distributed learning operation in real use cases, including the predictive performance, the associated communication overheads, or the energy consumption. In particular, we prove that distributed learning can improve the predictive accuracy centralized \gls{ml} solutions by up to~93\% while reducing the communication overheads and the energy cost by~80\%.
\end{abstract}

%% file: sections/1_introduction.tex
\section{Introduction}
\label{section:introduction}

In the rapidly evolving landscape of wireless communications, the accurate prediction of network metrics such as the load plays a crucial role in managing the network and optimizing performance. In IEEE 802.11 (aka Wi-Fi) deployments, \gls{ap} load prediction is becoming increasingly important for enabling proactive management and advanced network troubleshooting, thus contributing to meeting the stringent performance requirements (e.g., low latency, reliability) of new applications (e.g., virtual/extended reality) and use cases (e.g., industrial Wi-Fi)~\cite{galati2023will}. However, as a result of the highly dynamic nature of network conditions and user behaviors, Wi-Fi technology poses unique challenges for regression problems such as load prediction. In this regard, Machine Learning (ML) has been identified as a key enabler for addressing the load prediction problem in Wi-Fi~\cite{wilhelmi2023ai}, but the deployment of \gls{ml} solutions in Wi-Fi networks still entails unresolved challenges.

Wi-Fi networks are typically operated independently (e.g., home Wi-Fi, enterprise Wi-Fi), thus having limited access to data for training \gls{ml} models able to generalize effectively. Accordingly, the adoption of \gls{ml} models in those settings can result in a long start-up time (in the order of weeks or months), as it requires collecting a sufficiently large amount of data before starting to generate accurate predictions. Moreover, the high computational cost associated with running \gls{ml}-related operations (e.g., large data storage, intensive data processing, model training, model inference) prevents deploying complex \gls{ml} models into last-mile equipment such as Wi-Fi \glspl{ap} or edge servers, thus requiring either running low-complex \gls{ml} solutions (e.g., tiny \gls{ml}) or making use of additional hardware infrastructure. This, of course, entails a trade-off between the achievable model performance and its cost.

Against this background, distributed \gls{ml} emerges as an appealing solution to boost collaboration among multiple parties (e.g., independently operated deployments) and leverage their collective power to produce robust and ready-to-deploy \gls{ml} models. Distributed \gls{ml} model training allows multiple participants (or clients) to exchange relevant information for collaboratively training a given \gls{ml} model, rather than sharing raw training data (as it would be required in centralized \gls{ml}), which allows preserving privacy (the actual training data is not shared), keeping the communication overheads low (only \gls{ml} parameters are shared), and alleviating the computation burden of individual devices.

In this paper, we explore the advantages and disadvantages of adopting distributed \gls{ml} training solutions for the Wi-Fi \gls{ap} load prediction use case, where a set of Wi-Fi deployments are orchestrated to train powerful collaborative models. More specifically, we study two prominent distributed \gls{ml} paradigms, namely Federated Learning (FL)~\cite{mcmahan2017communication} and Knowledge Distillation (KD)~\cite{hinton2015distilling}, when applied to a dataset with real Wi-Fi measurements~\cite{chen2021flag}. The evaluation includes model performance, communication cost, memory footprint, and energy consumption, which are the pillars to determine the feasibility of adopting \gls{ml} solutions in practice. The distributed learning solutions are compared against \gls{cl}, whereby a single \gls{ml} model is trained by aggregating data from a collection of Wi-Fi \glspl{ap}.  

The rest of the document is structured as follows. Section~\ref{section:related_work} summarizes the related work on load prediction in wireless networks and on the application of distributed \gls{ml} in telecommunications problems. Section~\ref{section:distributed_learning} formulates the load prediction problem and describes the distributed training mechanisms adopted in this paper. Then, Section~\ref{section:methodology} details the experimental setup, which is followed by the evaluation results in Section~\ref{section:results}. The paper is concluded with final remarks in Section~\ref{section:conclusions}.

%% file: sections/2_related_work.tex
\section{Related Work}
\label{section:related_work}


In recent years, the prevalence of Wi-Fi network deployments and the proliferation of operator-managed solutions have fostered the emergence of \gls{ai} and \gls{ml} solutions for enhancing the management and operation of Wi-Fi networks~\cite{szott2022wi}. For the load prediction use case, limited related work can be found for the Wi-Fi domain, whereas a plethora of research works have been undertaken for cellular networks (we refer the interested reader to the survey in~\cite{jiang2022cellular}). In contrast to cellular technology, Wi-Fi's usage is mostly indoors, thus leading to substantially different behavioral aspects~\cite{cisco2019cisco}. Moreover, Wi-Fi operates in a distributed manner, which contrasts with the scheduled approach adopted in cellular.

With the highest degree of relation to this paper, the work in~\cite{wilhelmi2023ai} investigated the practical application of \gls{ml} for load prediction in enterprise Wi-Fi networks, emphasizing the potential benefits for autonomous operation and improved troubleshooting. The study shows that even in hardware-constrained environments, \gls{ml} models such as \gls{cnn} and \gls{lstm} achieve promising results with \gls{mape} results below \qty{20}{\percent} (in average) and below \qty{3}{\percent} (\num{85}-th percentile). These findings underscore the feasibility of leveraging \gls{ml} for proactive Wi-Fi network optimization, enhancing energy efficiency and performance. However, the proposed approach assumes a centralized \gls{ml} operation, whereby a single model is trained using the data from all the deployments and then applied to those deployments for performing inference. \gls{cl} was also studied in~\cite{chen2021flag}, where a novel \gls{rnn} architecture was proposed to address the load prediction problem in Wi-Fi. Such an \gls{rnn}-based solution was shown to improve the \gls{mape} achieved by state-of-the-art models like \gls{lstm} and \gls{arima} by \qty{25}\% and \qty{100}\%, respectively. In this paper, in contrast, we aim to explore the potentials and pitfalls of distributed learning.

When it comes to distributed \gls{ml} solutions for wireless networks, the work in~\cite{chen2021distributed} surveyed relevant state-of-the-art papers addressing challenges in terms of resource constraints, delay limitations, and privacy. However, as acknowledged in~\cite{chen2021distributed}, regression problems such as for load prediction are little studied. Moreover, practical considerations related to computation and communication requirements are rarely analyzed. For traffic prediction and related problems, the application of \gls{fl} approaches has gained certain popularity in the last years. \cite{liu2020privacy} applied \gls{fedavg} to solve traffic flow prediction across multiple organizations. A different approach was considered in~\cite{zhang2021dual}, which proposed a hierarchical FL approach for wireless traffic prediction, whereby similar \gls{bs} (grouped using K-means) participate in a tiered \gls{fl} operation. In~\cite{perifanis2023federated}, a comprehensive analysis and evaluation of multiple \gls{ml} models (e.g., \gls{rnn}, \gls{cnn}, \gls{lstm}) was conducted under various federated settings (e.g., \gls{fedavg}, FedProx, FedNova). A follow-up study of~\cite{perifanis2023federated} was performed in~\cite{perifanis2023towards}, where more advanced \gls{ml} model architectures like Transformer were considered. Furthermore, the evaluation done in~\cite{perifanis2023towards} provided relevant insights into the environmental sustainability of the studied models. Similarly, the work in~\cite{guerra2023cost} evaluated the energy consumption and the communication overheads of different distributed approaches, including \gls{fl} and gossip learning.

So far, most of the efforts in the literature have focused on solutions based on \gls{fl}. \gls{fl}, however, exhibits poor performance when working with non-\gls{iid} data~\cite{zhu2021federated}, which can be the case for independently managed Wi-Fi deployments willing to cooperate, which can be highly heterogeneous in terms of load. Therefore, more sophisticated solutions are much needed. One potential solution is \gls{kd}~\cite{hinton2015distilling}, which allows transferring knowledge from a complex or ensemble model $\theta_T$ (referred to as the \textit{teacher}) to a smaller (lightweight) model $\theta_S$ (referred to as the \textit{student}). In a federated setting, the clients act as the teacher and distill their knowledge into the server's (student) global model. To address the issues of \gls{fl} caused by user heterogeneity, the seminal work conducted in~\cite{zhu2021data} introduced a data-free \gls{kd} approach, where a lightweight generator \gls{ai} model is used to ensemble the clients' knowledge. The model is broadcasted to users, guiding local model learning as an inductive bias. The approach demonstrates improved generalization performance in \gls{fl} with fewer communication rounds compared to current state-of-the-art methods. Unfortunately, the approach in~\cite{zhu2021data} was primarily conceived for classification tasks, thus lacking applicability in regression problems.

In this paper, we contribute to the development and analysis of distributed \gls{ml} approaches for network traffic prediction, thus aiming at designing robust and efficient solutions for next-generation networks. In contrast to existing approaches, we not only focus on standard \gls{fl} techniques, but we also adopt the data-free \gls{kd} approach from~\cite{zhu2021data} (referred to as KD-gen in this paper) and extend it for the Wi-Fi \gls{ap} load prediction problem. In addition, we provide a holistic analysis of the studied solutions in terms of deployment feasibility, thus considering practical aspects including computation, communication, and energy consumption needs. 




%% file: sections/3_distributed_learning.tex
\section{System Model}
\label{section:distributed_learning}

\begin{figure*}[t!]
\centering
 \includegraphics[width=0.85\textwidth]{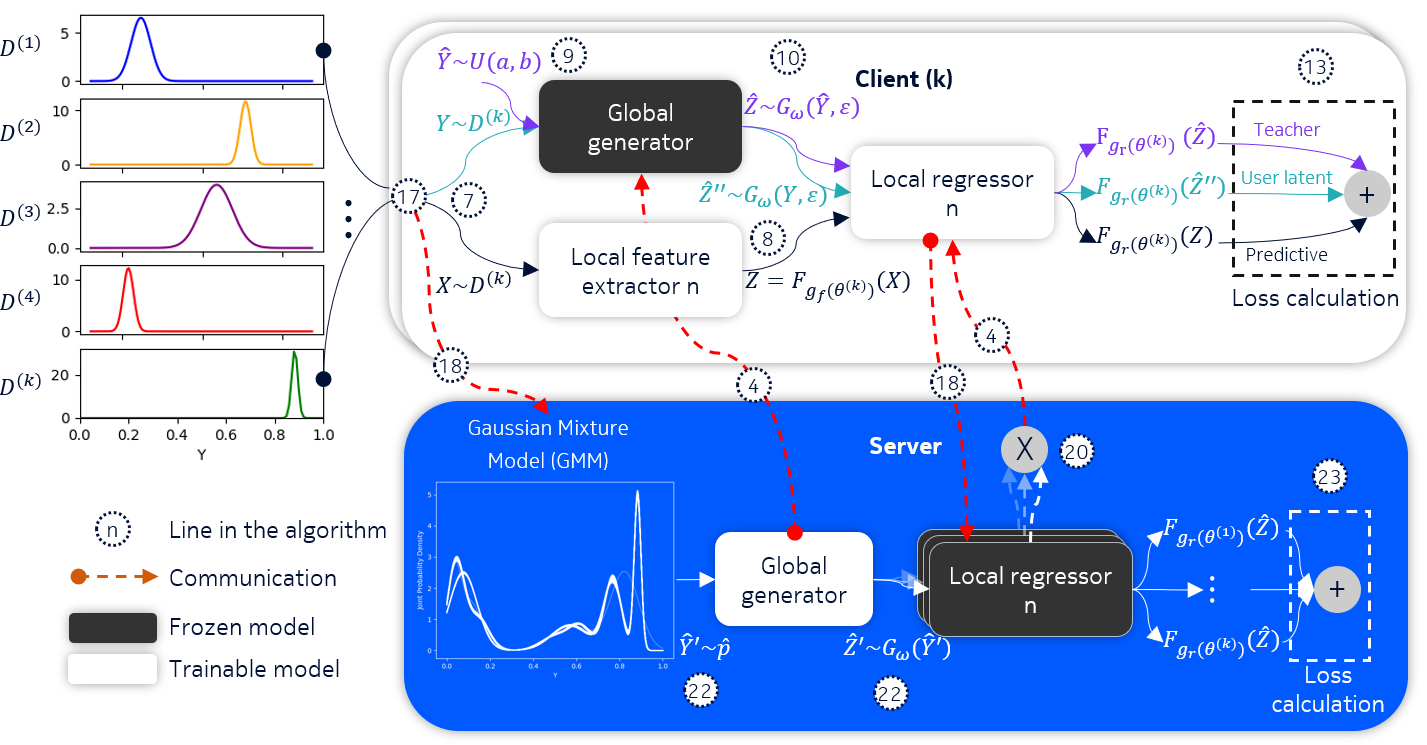}
 \caption{KD-gen detailed operation. The numbered bullets are aligned with the steps described in Algorithm~\ref{alg:distributed_training}.}
 \label{fig:proposed_scheme}
\end{figure*}

\subsection{ML-based Wi-Fi Load Prediction}

We target a regression problem using supervised learning to predict the load that a given Wi-Fi AP may experience in the future. To such end, AP $k$'s load and other relevant features are transformed into time-series signals $\mathbf{X}^{(k)}_{T} = \mathbf{x}^{(k)}_{1},\cdots,\mathbf{x}^{(k)}_{T}$, where $\mathbf{x}^{(k)}_t \in \mathbb{R}^d$. At a given time step $t$, the goal is to predict the load at future steps $s$ by using a lookback window of size $l$, i.e., $\mathbf{Y}^{(k)}_{t+1:t+s} = f(\mathbf{X}^{(k)}_{t-l-1:t})$. We use a neural network model to characterize the predictive function $f(\cdot)$. The weights of the neural network model, $\theta\in \mathbb{R}^d$, are derived by following different strategies, including \gls{cl}, \gls{fl}, and KD-gen. For the \gls{cl} case, which is used as a baseline, we assume that a global loss function $l(\cdot)$ is used to optimize the model weights $\theta$ of a global model so that $\theta^* = \min_{\theta} l(\theta)$. To approximate $\theta^*$, a given optimizer $o$ is used to minimize the losses observed on a given dataset $\mathcal{D}$. FL and KD-gen approaches extend the \gls{cl} baseline by assuming that the dataset $\mathcal{D}$ is distributed across several participants, thus not being available at a single location.

\subsection{Distributed Training}
\label{sec:training_models}

To train an ML model $\theta$ distributively, we assume a set of $\mathcal{K}=\{1,2,...,K\}$ clients that collaborate by exchanging relevant information but without exchanging their datasets directly. Each client holds a portion $\mathcal{D}^{(k)}$ with $D^{(k)}=|\mathcal{D}^{(k)}|$ of the dataset $\mathcal{D}$. To train a collaborative model, we use FL and KD-gen as distributed training mechanisms, which are evaluated against the centralized upper-bound (data is gathered at a single location) and the isolated lower-bound (each AP maintains its own model, thus no data is exchanged) in section~\ref{section:results}. Both FL and KD-gen are schematically illustrated in Fig.~\ref{fig:proposed_scheme} and algorithmically described in Algorithm~\ref{alg:distributed_training}. In both cases, the unique procedures from FL and KD-gen are represented in \ctext[RGB]{54,210,194}{turquoise} and \ctext[RGB]{153,204,255}{blue}, respectively, while the common procedures are not highlighted.


\begin{algorithm}[ht!]
	\caption{Distributed ML model training: \ctext[RGB]{54,210,194}{FL}, \ctext[RGB]{153,204,255}{KD-gen}.}\label{alg:distributed_training}
	\begin{algorithmic}[1]
	    \State \textbf{Initialize:} Initial model $\theta_0$, generative model \ctext[RGB]{153,204,255}{$\omega$}, batch size $B$, number of epochs $E$, learning rates $\eta$, \ctext[RGB]{153,204,255}{$\alpha$}, initial global distribution \ctext[RGB]{153,204,255}{$\hat{p}_0$}, random noise \ctext[RGB]{153,204,255}{$\epsilon \sim \mathcal{N}(0,1)$}.
		\For{$t=0,\ldots,T$}
		\State Select $\mathcal{S}_t \subseteq \mathcal{K}$ uniformly at random.
        \State Push $\theta_t$ and \ctext[RGB]{153,204,255}{$\omega_t$} to $\mathcal{S}_t$. 
		\For{$k \in S_{t}$}
    		\State Update the local model: $\boldsymbol{\theta}^{(k)}_{t,0} \leftarrow \boldsymbol{\theta}_{t}$.
            \State Sample mini-batch: $\{\mathbf{X}, \mathbf{Y}\}_1^B \sim \mathcal{D}^{(k)}$.
            \State Generate input features: \ctext[RGB]{153,204,255}{$\{\mathbf{Z} \sim F_{g_f({\theta}_{t,E}^{(k)})}(\mathbf{X})\}_1^B$}.
            \State Generate output from uniform distr.: \ctext[RGB]{153,204,255}{$\{\hat{\mathbf{Y}}\}_1^B \sim U$}.
            \State Generate induced features: \ctext[RGB]{153,204,255}{$\{\hat{\mathbf{Z}} \sim G_\omega(\hat{\mathbf{Y}}, \epsilon)\}_1^B$}.
    		\For{$e=1,\ldots,E$}
                \State \ctext[RGB]{54,210,194}{${\theta}^{(k)}_{t,e} \leftarrow {\theta}_{t,e}^{(k)}-\eta\nabla l(\theta^{(k)}_{t,e}, \{\mathbf{X},\mathbf{Y}\})$}.
                \State \ctext[RGB]{153,204,255}{${\theta}^{(k)}_{t,e} \leftarrow {\theta}_{t,e}^{(k)}-\eta\nabla l'(\theta^{(k)}_{t,e}, \{\mathbf{Z}, \mathbf{Y}\}, \{\hat{\mathbf{Z}}, \hat{\mathbf{Y}}\})$}. 
		\EndFor
		\State \ctext[RGB]{54,210,194}{${\theta}_{t+1}^{(k)} \leftarrow {\theta}_{t,E}^{(k)}$}.
        \State \ctext[RGB]{153,204,255}{${\theta}_{t+1}^{(k)} \leftarrow g_r({\theta}_{t,E}^{(k)})$}.    
        \State Update local distribution: \ctext[RGB]{153,204,255}{$p^{(k)}_{t+1} \leftarrow e(\mathbf{X},\mathbf{Y})$}.
        \State Push $\theta^{(k)}_{t+1}$ and \ctext[RGB]{153,204,255}{$p^{(k)}_{t+1}$} to the server. 
		\EndFor
		\State Model aggregation: ${\theta}_{t+1} \leftarrow \frac{1}{ \vert \mathcal{S}_t \vert}\sum_{k\in \mathcal{S}_t} {\theta}_{t+1}^{(k)}$. 
        \State Update global distr.: \ctext[RGB]{153,204,255}{$\hat{p}_{t+1} \leftarrow \text{GMM}(\{p^{(k)}_{t+1}\}_{k\in S_t})$}.
        \State \ctext[RGB]{153,204,255}{Generate $\{\hat{\mathbf{Z}}', \hat{\mathbf{Y}}'\}$ from $G_\omega$ and $\hat{p}$}.
        \State Update gen. model: \ctext[RGB]{153,204,255}{${\omega}_{t+1} \leftarrow {\omega}_t -\alpha\nabla l({\omega}_t, \{\hat{\mathbf{Z}}', \hat{\mathbf{Y}}'\})$}.
		\EndFor
	\end{algorithmic}
\end{algorithm}

\subsubsection{Federated learning}

\gls{fl}~\cite{mcmahan2017communication} is a decentralized \gls{ml} paradigm whereby an orchestrating server aggregates the individual model weights retrieved from a set of participants $\mathcal{K}$ in an iterative manner, for $t = 0,...,T$ \gls{fl} rounds. In a given \gls{fl} round $t$, the server pushes a global model $\theta_t$ to a subset of clients $S_t \subseteq \mathcal{K}$ (line \num{5} in Algorithm~\ref{alg:distributed_training}), lets the participants $k\in S_t$ train the model $\theta_t$ locally by running $E$ epochs on their local datasets $\mathcal{D}^{(k)}$ and computed loss $l^{(k)}_t$ (lines \numrange{10}{13} in Algorithm~\ref{alg:distributed_training}), and finally retrieves the resulting local models $\theta^{(k)}_{t+1}, \forall k\in S_t$ to update the model through aggregation (line \num{19} in Algorithm~\ref{alg:distributed_training}). Based on Federated Averaging (FedAvg)~\cite{mcmahan2017communication}, the global model is empirically optimized by minimizing the loss expectation across the clients' losses, i.e., $\theta^* = \min_{\theta} \frac{1}{K} \sum_{k=1}^K l(\theta, \mathcal{D}^{(k)})$.


\subsubsection{Knowledge distillation with generative model}

\gls{kd}'s primary goal is to distill the generalization capabilities of the teacher model $\theta_T$ into the more computationally efficient student model $\theta_S$. This is typically done by minimizing the discrepancy between the teacher and the student models' last layer on a public proxy dataset~\cite{chen2021distributed}, so that $\theta_S^*  = \min \rho(\sigma(\theta_T) || \sigma(\theta_S))$, where $\rho$ is a distance metric like the Kullback-Leibler (KL) divergence and $\sigma$ is the function that retrieves the normalized weights of the model's last layer. The problem with this approach is that it requires a powerful and publicly available dataset to be used as a reference. To bypass such a strong requirement, \cite{zhu2021data} proposes KD-gen, through which a generative model is trained and exchanged for the sake of distilling the clients' knowledge.

KD-gen is applied to a distributed training setting so that the knowledge of $K$ clients (teachers) is transferred into a parameter server (student), which aggregates and sends it back to clients iteratively. Following the approach in~\cite{zhu2021data}, this is achieved by training and sharing a generator $G_\omega$ parameterized by $\omega$, through which clients $k\in \mathcal{K}$ generate synthetic samples $\hat{\mathbf{Z}}$ (from a latent feature space $\mathcal{Z} \in \mathbb{R}^d$) that are used for training the local models $\theta^{(k)}$. More specifically, at a given training round $t$, clients sample prediction outputs $\hat{\mathbf{Y}}$ from $U$ which is a uniform distribution (Line \num{9} in Alg.~\ref{alg:distributed_training}) and use them as inputs for the generator $G_\omega$, which in turn creates synthetic features $\hat{\mathbf{Z}}$ based on the generative model's weights $\omega_t$ (Line \num{10} in Alg.~\ref{alg:distributed_training}).

The generative model is trained on the orchestrating server using the empirical representation of all the clients' data $\hat{p}$, obtained as a mixture of the clients' data distributions, i.e., $\hat{p} = \text{GMM}(\{\hat{p}^{(k)}\}_{k\in \mathcal{K}})$. A loss function $l'(\cdot)$ is used to optimize the local model $\theta^{(k)}_t$, so $l'(\cdot)$ includes the losses for both real and synthetic samples, i.e., $\{\mathbf{Z}, \mathbf{Y}\}$ and $\{\mathbf{\hat{Z}}, \mathbf{\hat{Y}}\}$ (Lines \numrange{11}{14} in Alg.~\ref{alg:distributed_training}). Note that, as shown in Line \num{8} in Alg.~\ref{alg:distributed_training}, the input features $\mathbf{Z}$ is calculated using a few beginning layers of the local model ($g_f(\theta^{(k)}_{t+1})$). Then, only a few layers of the resulting model $\theta'^{(k)}_{t+1} = g_r(\theta^{(k)}_{t+1})$ are passed to server together with the updated local data distribution $p^{(k)}_{t+1}$ (Line \num{18} in Alg.~\ref{alg:distributed_training}). Finally, the server aggregates the models and data distributions received from the clients to generate $\theta_{t+1}$ and $\hat{p}_{t+1}$, respectively, and updates the generative model $\omega_{t+1}$ (Lines \numrange{20}{22} in Alg.~\ref{alg:distributed_training}). For this last step, synthetically generated data $\{\hat{\mathbf{Z}}', \hat{\mathbf{Y}}'\}$ from the \gls{gmm} model and the generator is used, to maximize the posterior probability $p(\hat{\mathbf{Z}}'|\hat{\mathbf{Y}}')$. 



%% file: sections/4_methodology.tex
\section{Experimental Setup \& Methodology}
\label{section:methodology}

In this section, we describe the experimental setup and evaluation scenario considered in this paper, for which Table~\ref{tab:parameters} provides relevant parameters.

\begin{table}[ht!]
\centering
\caption{ML model and evaluation parameters.}
\label{tab:parameters}
\resizebox{\columnwidth}{!}{%
\begin{tabular}{lc}
\hline
\textbf{Parameter} & \multicolumn{1}{c}{\textbf{Value}} \\ \hline
Number of users, $K$ & 100 \\
Number of deployments, $U$ & 20 \\
Number of users per round, $S$ & 20 \\
Train split ratio, $\rho_\text{train}$ & 0.8 \\
Test split ratio, $\rho_\text{test}$ & 0.2 \\
Lookback window, $l$ & 60 \\
Prediction step, $s$ & \{1, 5, 15, 30\} \\
Number of global epochs (CL), $E_g$ & 100 \\
Number of local epochs (FL and KD-gen), $E_l$ & 50 \\
Distr. learning rounds (FL and KD-gen), $R$ & 100 \\
LSTM/FC layers, $N_\text{lstm}/N_\text{fc}$ & 1/2 \\
LSTM hidden layer size, $H$ & 480 \\
Loss function, $l$ & L1 \\
Optimizer, $o$ & Adam \\
Learning rate, $\eta$ & 0.01 \\
Batch size, $B$ & 32 \\
Dataset size, $|\mathcal{D}|$ & 750 MB \\ \bottomrule
\end{tabular}%
}
\end{table}

\subsection{Data}

\begin{figure*}[ht!]
\centering
 \includegraphics[width=0.9\textwidth]{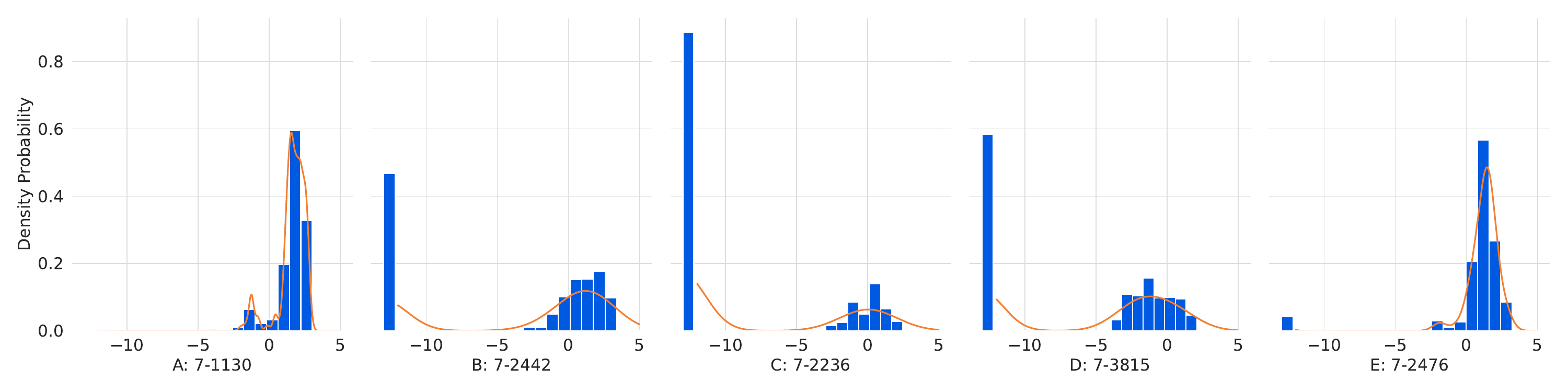}
 \caption{Histograms and \gls{kde} curves for load of five different \glspl{ap}}
 \label{fig:histograms_and_kde}
\end{figure*}

The evaluation of the distributed \gls{ml} solutions is done on the open-source dataset disclosed in~\cite{chen2021flag}, which contains measurements collected from \num{7404} Wi-Fi \glspl{ap} during \num{49} days. As done in~\cite{wilhelmi2023ai}, we derive the load of each \gls{ap} and downsample it in windows of $W = 2$ minutes. The load of a given \gls{ap} is transformed along with other available features (i.e., number of users connected, hour of the day, and day of the week) into time series arrays $\mathbf{X} = \{\mathbf{X}_{l}, \mathbf{Y}_{s}\}$, where feature ($\mathbf{X}_{l}$) and label ($\mathbf{Y}_{s}$) arrays are created using a sliding window, so that $\mathbf{X}_{l}^{(i)} \in \mathbb{R}^{d\times l}$ and $\mathbf{Y}_{s}^{(i)} \in \mathbb{R}^{1\times s}$, for each $i\in\mathbf{X}$. Data is normalized using a Min-Max scaler so that it is bounded by the range $[1,2]$. An excerpt of the dataset distribution is illustrated in Fig.~\ref{fig:histograms_and_kde}, which includes the histograms of the normalized load from five different \glspl{ap} in the dataset.

\subsection{Evaluation Scenario}

We consider $\mathcal{U}$ (with $|\mathcal{U}|=20$) different deployments containing \num{4}-\num{6} \glspl{ap}, which are obtained by applying a Dirichlet on a set of $K=100$ \glspl{ap} randomly selected from the dataset described above. The time series data of each \gls{ap} is split into train ($\mathbf{X}_\text{train}$) and test ($\mathbf{X}_\text{test}$) sets, based on split ratios $\rho_\text{train}$ and $\rho_\text{test} = 1-\rho_\text{train}$. Depending on the training approach and the available data, \gls{ml} model training and evaluation are performed differently. In particular:
\begin{itemize}
    \item In \textbf{Isolated \gls{cl} (ICL)}, each considered deployment is treated independently, so that a different model is maintained per deployment. Accordingly, the model of a specific deployment is trained and tested using only the data from the \glspl{ap} within the same deployment.
    \item In \textbf{Data-Sharing \gls{cl} (DSCL)}, a single model is trained and evaluated by using the data from all the \glspl{ap} across all the considered deployments.
    \item In the distributed learning settings (\textbf{FL} and \textbf{KD-gen}), a local model is maintained for each deployment (thus using only local data for training), but a single global model is generated based on parameter exchanges. The resulting global model is evaluated on each deployment.
\end{itemize}


 
\subsection{ML Model}

The \gls{ml} model $f(\cdot)$ used to predict the load is \gls{lstm}, which combines \gls{lstm} (including \gls{lstm} cells) and linear layers to capture temporal correlations and generate future predictions, respectively. The election of \gls{lstm} is based on previous works showing its suitability for the predictive task at hand~\cite{perifanis2023federated}. In particular, the considered \gls{lstm} model includes an input layer of size ($B$, $d$, $l$), $N_\text{lstm}$ \gls{lstm} layers of size $H$, and $N_\text{fc}$ fully-connected layers that output $\max(s)$ predictions. The resulting \gls{lstm} model has \num{953342} trainable parameters (i.e., $|\theta| = 3.7$~MB and $|\theta'|=2\cdot10^{-3}$~MB for float64 format), which leads to \num{3.61} \gls{gmac}. In addition to the LSTM, the KD-gen approach uses a generative model $\omega$, which has \num{26368} parameters, thus requiring $|\omega| = 0.113$~MB of memory (corresponding to \num{1.7}~\gls{gmac}).




\subsection{Performance: Accuracy, Overheads, Energy, and Time}

The performance evaluation of the models does not only include predictive accuracy, but also practical aspects, including communication overheads, energy spent, and computation time. More specifically, we define the evaluation performance metrics as follows:

\begin{itemize}
    \item \textbf{Predictive performance, $\zeta$ [MB]:} The predictive performance of the ML models is assessed using the \gls{mae}, which is defined as follows:
    \begin{equation}
        \text{MAE} = \frac{\sum_{\mathbf{y}\in \mathbf{Y}} \lvert \hat{\mathbf{y}} - \mathbf{y} \rvert}{\lvert \mathbf{Y} \rvert} .
    \end{equation}
    \item \textbf{Communication overhead, $C$ [MB]:} The communication cost of each learning setting is computed as follows:\footnote{For KD-gen, the size of the distribution parameters $p$ is negligible compared to the rest of the exchanged parameters, so it is excluded from the analysis.}
    \begin{equation}
    \resizebox{0.8\hsize}{!}{%
        $C(x)=
        \begin{cases}
            \sum_{k\in K} \big(D^{(k)} + |\theta| \big), & \text{if } x = \text{CL}, \\
            2 |\theta| \sum_{t=0}^{T-1}|S_t|,  & \text{if } x = \text{FL}, \\
            \big(2 |g_r(\theta)| + |\omega|\big) \sum_{t=0}^{T-1}|S_t|, & \text{if } x = \text{KD-gen}, \\
            0, & \text{if } x = \text{IL}.
        \end{cases}$
    }
    \end{equation}
    
    %
    \item \textbf{Energy consumption, $\sigma$ [Wh and CO$_2$ kg]:} The energy spent on training and inference tasks is computed using \textit{eco2AI}'s Python library~\cite{budennyy2022eco2ai}, which estimates GPU, CPU, and RAM energy consumption through hardware measurements.
    \item \textbf{Training time, $T_\text{train}$ [s]:} The wall-clock time required to train each solution.
\end{itemize}

%% file: sections/5_results.tex
\section{Results and Discussion}
\label{section:results}

\begin{figure}[ht!]
    \centering
     \includegraphics[width=.9\columnwidth]{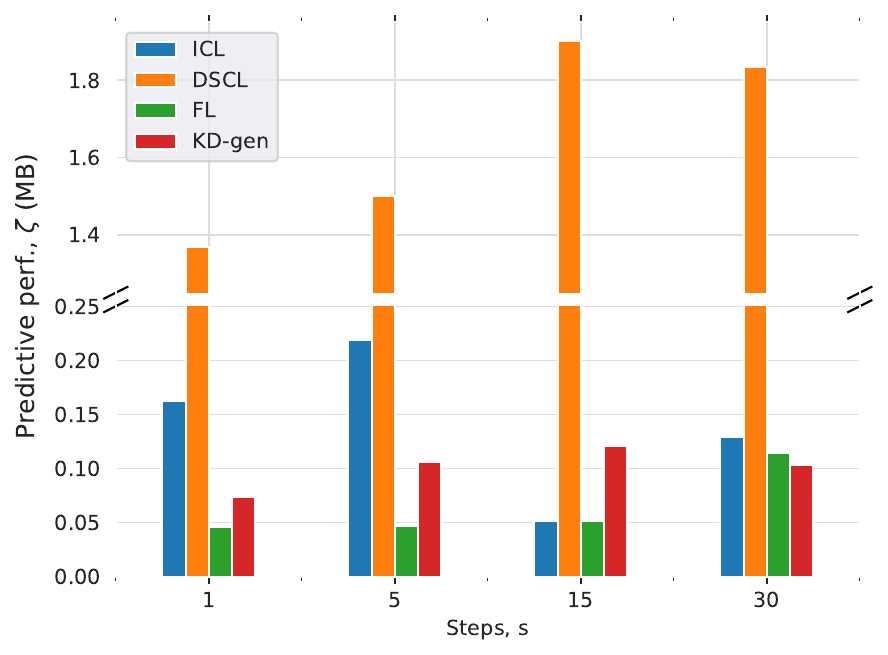}
     \caption{\gls{mae} of \{ICL, DSCL, FL, KD-gen\} at $s\in\{1,5,15,30\}$.}
     \label{fig:approach_comparison}
\end{figure}

We start showing the predictive performance $\zeta$ achieved by the considered approaches in Fig.~\ref{fig:approach_comparison}, where the mean \gls{mae} (in MB) achieved across all the deployments is shown for each considered prediction step $s\in\{1,5,15,30\}$. 

As it can be observed in Fig.~\ref{fig:approach_comparison}, the distributed learning approaches (\gls{fl} and KD-gen) outperform the \gls{cl} baselines (ICL and DSCL) in most cases, thereby confirming both \gls{fl} and KD-gen as appealing solutions to extract and transfer useful knowledge from the different cooperating deployments. More specifically, we observe that \gls{fl} provides better results than KD-gen for $s = \{1,5,15\}$, which is due to the data IIDness that stems from the high resemblance among the different deployments. For $s=30$ (i.e., predictions \num{1}~hour ahead in the future), however, KD-gen shows slightly better performance than \gls{fl}. This reveals the ability of KD-gen to address far-distant prediction horizons, where IIDness properties are broken due to the increasing uncertainty. Uncertainty is precisely what motivates the collaboration among multiple deployments through distributed learning so that the collective knowledge can be leveraged to anticipate complex and challenging events. Finally, for DSCL in particular, we observe a significantly higher error compared to the rest of the approaches, which is motivated by the fact that a single model, even if trained on a large dataset comprising measurements from $N=100$ \glspl{ap}, does not fit all the deployments. As for ICL, it offers a better performance than DSCL thanks to the better specialization that is achieved for each of the deployments.

To complement the abovementioned results, we now provide insights on the practical cost of training the \gls{ml} models through each of the considered approaches. In particular, Table~\ref{tab:cost_approaches} provides the communication overheads, energy spent (both in Wh and CO$_2$~kg), and the clock-wall training time associated with ICL, DSCL, FL, and KD-gen. 

\begin{table}[ht!]
\centering
\caption{Practical cost of training the ML models with each approach.}
\label{tab:cost_approaches}
\resizebox{\columnwidth}{!}{%
\begin{tabular}{@{}ccccc@{}}
\toprule
 & \textbf{ICL} & \textbf{DSCL} & \textbf{FL} & \textbf{KD-gen} \\ \midrule
$C$ (MB) & 0$^b$ & 1120 & 14800 & 234 \\
$\sigma$ (Wh)$^a$ & 9.69 & 411.58 & 52.53 & 81.97 \\ 
$\sigma$ ($\text{CO}_2$ kg)$^a$ &  $1.3\cdot10^{-3}$ & $58.4\cdot10^{-3}$ & $7.45\cdot10^{-3}$ & $11.63\cdot10^{-3}$ \\ 
$T_\text{train}$ (s)$^a$ & 195.47 & 6665.82 & 594.39 & 958.41 \\ \bottomrule
\vspace{0.1em}
\end{tabular}%
}
\raggedright
{\footnotesize $^a$The evaluation is carried out in an AMD Ryzen Threadripper 3970X 32-Core Processor machine.}\\
{\footnotesize $^b$Intra-deployment communication (e.g., from APs to the deployment's controller) is not considered.}
\end{table}

Starting with the communication requirements ($C$), ICL does not incur any overheads, provided that data is handled internally in each deployment. DSCL, in contrast, leads to significant overheads (\num{1120} MB) for completing the task at hand, as the training data needs to be transferred to a central point. Notice, as well, that in this case, the dataset was fairly small (only a few features were considered), while more challenging data (e.g., including images) would dramatically increase the communication needs for DSCL. As for \gls{fl}, it leads to the highest overheads (\num{14800} MB), which results from the numerous times the entire model needs to be pushed to and from the server during the training phase. Finally, KD-gen leads to moderate overheads (\num{234} MB), hence standing out as a sustainable option. 

In terms of energy consumption ($\sigma$), ICL and DSCL are the lowest and the highest consuming approaches with \num{9.69}~Wh and \num{411.58}~Wh, respectively. FL and KD-gen lead to a similar consumption (\num{52.53}~Wh and \num{81.97}~Wh), being KD-gen's bigger as a result of its additional operations related to the generative model. Finally, the training time ($T_\text{train}$) shows a similar trend to energy, as both metrics are strongly correlated. It is important to highlight that, both FL and KD-gen decrease one order of magnitude the time required by the DSCL approach, where the global model becomes a bottleneck.

%% file: sections/6_conclusions.tex
\section{Conclusions}
\label{section:conclusions}

Distributed learning emerges as a promising paradigm to make \gls{ml} suitable in practical use cases, as it allows addressing important concerns such as \gls{ml} model generalization, communication cost, energy consumption, and security. In this paper, we have studied two trendy distributed learning paradigms such as \gls{fl} and \gls{kd} when applied to telecommunications. More specifically, we have targeted the load prediction problem in enterprise Wi-Fi \glspl{ap}, thus aiming to illustrate the benefits that collaboration through distributed learning may bring to such a setting. Our results, obtained using real data measurements from a large Wi-Fi campus network, have demonstrated the great advantages of distributed learning against standalone and centralized \gls{ml} model training. More specifically, we have shown that FL and KD-gen outperform the isolated and central approach in terms of predictive performance, thus showcasing the need for cooperating. Furthermore, we have shown that KD-gen offers an excellent compromise between predictive accuracy and deployment cost, thus standing out as a promising solution to be further explored.